\documentclass[12pt]{iopart}
\usepackage{graphicx}
\usepackage{amssymb}
\usepackage{lscape}
\begin{document}

\title{Size quantization effects in thin film Casimir interaction}
\date{\today}

\author{A.~Benassi$^{1,2}$, C.~Calandra$^2$}

\address{$^1$ CNR/INFM-National Research Center on nanoStructures and bioSystems at Surfaces
(S3), Via Campi 213/A, I-41100 Modena, Italy}
\address{$^2$ Dipartimento di Fisica, Universit\`a di Modena e Reggio Emilia, Via Campi 213/A, I-41100 Modena,
Italy} \ead{benassi.andrea@unimore.it}

\begin{abstract}
We investigate the role of size quantization in the vacuum force between metallic films of nanometric thickness. The force is calculated by the 
Lifshitz formula with the film dielectric tensor derived from the one-electron energies and wavefunctions under the assumption of a constant 
potential inside the film and a uniform distribution of the positive ion charge. The results show that quantization effects tend to reduce the force 
with respect to the continuum plasma model. The reduction is more significant at low electron densities and for film size of the order of few nanometers 
and persists for separation distances up to $10\div 50$ nm. Comparison with previous work indicates that the softening of the boundary potential is 
important in determining the amount of the reduction. The calculations are extended to treat Drude intraband absorption. It is shown 
that the inclusion of relaxation time enhances the size quantization
effects in the force calculations.
\end{abstract}

\pacs{}

\section{Introduction}
Thin metal films of nanometric size can have physical properties different from their bulk counterparts due to quantum size effects (QSE) caused by 
the valence electron confinement in the direction normal to the film surfaces \cite{chiang,tringides,jia}. Since thin films are among the basic 
components of modern nanodevices an increasing attention has been addressed to the quantum vacuum fluctuations forces (van der Waals and Casimir 
forces) between them. Theoretical determinations of these forces have been based mainly on a continuum description of the material dielectric 
properties, neglecting modifications of the electronic structures due to the boundaries, i.e. under the assumption that the film dielectric 
properties be the same of the bulk material \cite{esquivel,benassi,lambrecht2,pirozhenko2,lenac}. This assumption is not valid for small systems with low electron 
density and/or large surface to volume ratio. In such cases a significant size dependence of the basic properties, like the Fermi energy and the 
density of electron states, arises as a consequence of the electron confinement \cite{rogers1}.\\
In a previous paper \cite{benassi3} we have shown that the inclusion of QSE in the plasma model of a free electron metal leads to modifications of the 
force intensity between nanometric size films with respect to the bulk plasma model which range from several to few percent depending upon the film 
electron density and the separation distance. The calculated forces show quantum size oscillations and are less intense compared 
to those determined with the bulk plasma model. The calculations were performed using the \emph{particle in a box} model \cite{wood} (hereafter indicated as PBM) in which independent electrons 
are confined along the direction normal to the film surfaces by hard walls and behave as a two dimensional gas parallel to the 
surfaces. Such model represents a simplified picture of the one electron potential in the film.\\
In this paper we improve our description of the film dielectric properties along two lines of development: 
first we introduce a finite well model for the one-electron potential along the surface normal, second we include intraband absorption by introducing the relaxation time in a manner that allows to keep number conservation.
As to the first issue, we notice that the use of a soft confining potential like a finite well, besides being closer to the real shape of the one-electron potential as determined by first 
principles calculations \cite{schulter,feibelman,ciraci,ogando}, allows for a better treatment of QSE for films of small thickness, since it does not 
introduce a priori a distance 
within which electrons are to be confined and leaves the Fermi energy free to oscillate with the film size \cite{rogers1,rogers2}, a feature that is 
not present in the PBM model, where the Fermi energy is kept equal to the bulk value. 
The inclusion of Drude absorption in vacuum force calculations has been a subject of considerable debate, mainly in relation with finite temperature 
corrections \cite{boestrom3,bezerra,brevik,bezerra2,hoye2,brevik2}. Very accurate measurements have been reported that shows that the simple plasma model of bulk dielectric can give better agreement with the experiments than the model with finite relaxation times \cite{decca2}.
Our results refer to the $T=0^{\circ}$K case where difficulties do not seem to be present \cite{lamoreaux}. Our purpose is to understand how QSE may affect the force between thin films when intraband absorption is included.
In this paper we introduce such corrections and we show their importance in the 
calculation of the intensity of the force between metallic films.\\
\section{Theoretical framework}
\begin{figure}
\centering
\includegraphics[width=12cm,angle=0]{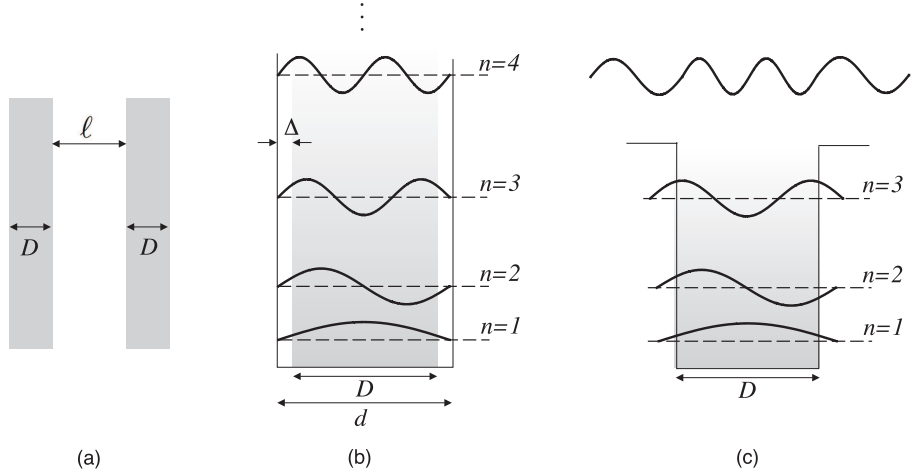}
\caption{\label{fig1} (a) Sketch of the two identical interacting films. (b) Infinitely deep quantum well with artificial spill-out. (c) Finite quantum well with natural spill-out.}
\end{figure}

We consider two identical metal films of thickness $D$ with plane boundaries separated by a distance $\ell$ (see figure \ref{fig1} (a)). $D$ represents 
the extension, along the $z$-direction normal to the surface, of the positive ion distribution, which is supposed to be uniform with the same density of the bulk system. By extending 
previous results for isotropic slabs to the case of films with anisotropic dielectric tensor we can write the expression of the force per unit area 
as \cite{lifshitz,dzyaloshinskii,zhou,bordag}:
\begin{equation}
F=-\frac{\hbar }{2 \pi^2}\int_{0}^{\infty} k\: dk\int_{0}^{\infty}d\omega\: \gamma(\omega)
\bigg[\frac{Q_{TM}(i\omega)^2}{1-Q_{TM}(i\omega)^2}+
+\frac{Q_{TE}(i\omega)^2}{1-Q_{TE}(i\omega)^2}\bigg] 
\end{equation}
\begin{equation}
Q_{TM}=\frac{\rho_{TM}(1-e^{-2 \gamma_{TM} D})}{1-\rho_{TM}^2e^{-2 \gamma_{TM} D}}e^{-\gamma \ell}\qquad
Q_{TE}=\frac{\rho_{TE}(1-e^{-2 \gamma_{TE} D})}{1-\rho_{TE}^2e^{-2 \gamma_{TE} D}}e^{-\gamma \ell}
\end{equation}
\begin{equation}
\rho_{TM}=\frac{\gamma_{TM}(\omega)-\gamma\epsilon_{xx}(\omega)}{\gamma_{TM}(\omega)+\gamma\epsilon_{xx}(\omega)}    
\qquad\rho_{TE}=\frac{\gamma_{TE}(\omega)-\gamma(\omega)}{\gamma_{TE}(\omega)+\gamma(\omega)}
\end{equation}
\begin{equation}
\gamma_{TE}(\omega)=\sqrt{k^2-\frac{\omega^2}{c^2}\epsilon_{xx}(\omega)}\qquad
\gamma_{TM}(\omega)=\sqrt{\bigg(\frac{k^2}{\epsilon_{zz}(\omega)}-\frac{\omega^2}{c^2}\bigg)\epsilon_{xx}(\omega)}
\end{equation}
\begin{equation}
\gamma(\omega)=\sqrt{k^2-\frac{\omega^2}{c^2}}
\end{equation}
Here $\epsilon_{xx}$ and $\epsilon_{zz}$ are the diagonal components of the dielectric tensor along the planar directions and along the surface normal respectively. We assume $\epsilon_{xx}=\epsilon_{yy}$ and the off diagonal components to be
zero. 
This assumption is consistent with the two dimensional gas behavior of the electrons parallel to the surface. The anisotropy of the dielectric tensor is a consequence of the finite extension of the film along the $z$-direction and it is the main feature introduced by the size quantization. To calculate the force one needs the expression of the dielectric tensor, which has to be derived from the film electronic structure. 
We assume an independent particle model with the one electron potential $V(z)$. The electron energies are given by
\begin{equation}
E_{\textbf{k}_{\parallel},n}=\frac{\hbar^2}{2 m}\textbf{k}_{\parallel}^2+E_{n}
\end{equation} 
i.e. they are described by the continuous quantum number $\textbf{k}_{\parallel}$ and by the discrete sub-band index $n$ coming from the quantization of the perpendicular wavevector, $m$ being the electron mass. The corresponding wavefunctions are given by
\begin{equation}
\psi_{\textbf{k}_{\parallel},n}(\textbf{r}_{\parallel},z)=\frac{1}{\sqrt{A}}e^{i\textbf{k}_{\parallel}\cdot\textbf{r}_{\parallel}}\phi_{n}(z)
\label{wfc}
\end{equation} 
here $A$ is the surface area, $\textbf{k}_{\parallel}$ and $\textbf{r}_{\parallel}$ are two dimensional wavevectors parallel to the surface, and $\phi_{n}(z)$ is supposed to be independently normalized. The functions 
$\phi_{n}(z)$ are the solutions of the equation
\begin{equation}
\bigg\{\frac{\hbar^2}{2 m}\frac{\partial^2}{\partial z^2}+V(z)\bigg\}\phi_{n}(z)=E_n\phi_{n}(z)
\label{schrod}
\end{equation}
with the proper boundary conditions. The Fermi energy $E_{F}$ is obtained through the \emph{aufbau} procedure i.e. by arranging the eigenvalues in 
ascending numerical order and counting until the number of states needed to accommodate all the electrons in the film is reached. This procedure 
leads to a Fermi energy that depends upon the film size and is generally different from the bulk value \cite{rogers2}. This can be understood 
by noting that, to ensure charge neutrality, the number of electrons and the number of ions per unit area have to be equal. The electron density 
$n(z)$ can be simply obtained from the wavefunctions:
\begin{equation}
n(z)=\frac{1}{2 \pi}\sum_{m=1}^{m_0}\bigg(\frac{2 m E_F}{\hbar^2}-E_n\bigg)\vert\phi_{m}(z)\vert^2
\end{equation} 
where $m_0$ is the label of the last occupied state, while the ion density is simply given by $n_0=k_{F_{B}}^3/3 \pi^2$, where $k_{F_{B}}$                                               is the bulk Fermi wavevector. By integrating the densities along the $z$ axis and imposing that both give the same number of charges per unit area, one gets the relation
\begin{equation}
\frac{1}{2 \pi}\sum_{m=1}^{m_0}\bigg(\frac{2 m E_F}{\hbar^2}-E_n\bigg)=n_0 D
\label{neucond}
\end{equation}
which, for finite $D$ values, is generally not satisfied if one replaces $E_F$ with its bulk counterpart $E_{F_{B}}=\hbar^2 k_{F_{B}}^2/2 m$. 
In the case of the PBM model this equation is not satisfied, since one assumes that the Fermi level is the same as in the bulk. To obtain charge neutrality one has to 
impose the additional condition that the electron density be confined on a length $d$ larger than $D$. This artificially introduces the electronic charges spill-out but has the consequence that the average 
electronic density is lower than the ion density \cite{rogers1,benassi3,czoschke,czoschke2}.\\  
For the purpose of the present study we assume the potential to be a finite well $V(z) = -V_0$ inside the film and zero outside. For such finite well
model (FWM) the energies of the bound states can be written as 
\begin{equation}
E_{\textbf{k}_{\parallel},n}=\frac{\hbar^2}{2 m}\textbf{k}_{\parallel}^2+\frac{\hbar^2}{2 m}k_{zn}^2-V_0
\end{equation}
where $k_{zn}$ are the quantized transverse wavevectors. They are obtained from the equation giving the condition for the existence of bound states 
in a quantum well \cite{davydov}:
\begin{equation}
k_{zn}=\frac{n \pi}{D}-\frac{2}{D}sin^{-1}\bigg(\frac{k_{zn}}{k_0}\bigg)
\label{russi}
\end{equation}
with $k_0=\sqrt{2 m V_0}/\hbar$.  Notice that the first term at the second member is the value of the transverse wavevector for an infinite well 
model of size $D$ and the second term goes to zero as $V_0$ goes to infinity. This implies that, for given film size and number of electrons, 
the Fermi energy referred to the well bottom is higher in the infinite well model.
Notice that when $V_0$ goes to infinity one does not recover the PBM, since the Fermi energy is varied with respect to the bulk value in order to 
satisfy the charge neutrality condition (\ref{neucond}). Figure \ref{fig1} (b) and (c) illustrate the difference: in the PBM the electronic charge density is 
confined within a distance $d=D+\Delta$, larger than the size of the ionic charge distribution, to allow for the electronic charge spill-out and to 
ensure global neutrality for $E_F=E_{F_{B}}$. In the FWM the charge spill-out results naturally from the behaviour of the single particle 
states while the charge neutrality is achieved by varying the Fermi energy with respect to the bulk value. Obviously in the limit of infinitely deep well 
(hereafter indicated as IWM) the electronic charge turns out to be entirely localized within the length $D$ and the Fermi energy is strongly 
increased compared to its bulk value.\\
Once the electron energies and wavefunctions have been obtained, one can calculate the dielectric tensor from the expression \cite{wood}:
\begin{eqnarray}
\nonumber
\epsilon_{\alpha\alpha}(\omega)&=1-\frac{\omega_{P}^{2}}{\omega^{2}}-\frac{8\pi e^{2}}{A d m^{2}
\omega^{2}}\sum_{{\bf k}_{\parallel},n}\sum_{{\bf k}_{\parallel}',n'}
f(E_{\bf{k}_{\parallel},n})(E_{\bf{k}_{\parallel},n}-E_{\bf{k}_{\parallel}',n'})\times\\
&\times\frac{\vert\langle\psi_{{\bf k}_{\parallel},n}\vert\hat{p_{\alpha}}\vert\psi_{{\bf k}_{\parallel}',n'}
\rangle\vert^{2}}{(E_{\bf{k}_{\parallel},n}-E_{\bf{k}_{\parallel}',n'})^2-\hbar^2\omega^2}
\label{dieltens}
\end{eqnarray}
here $\alpha=x,y,z$ labels the cartesian component of the tensor, $\hat{p}_{\alpha}$ indicates the component of the electron linear momentum,
$\omega_{P}=\Omega_{P} n(D)/n_0$ is the plasma frequency of the quantized electron gas ($\Omega_{P}=\sqrt{4\pi e^2 N_0/m}$ is the free electron plasma frequency and $n(D)$ is the average electron density of the film) and $f(E_{\bf{k}_{\parallel},n})$ is the occupation factor of the $(\bf{k}_{\parallel},n)$ state. 
In the IWM and the FWM $\omega_P = \Omega_P$, while in PBM $n(D)$ is smaller than $n_0$, since the electronic charge is distributed over a larger distance than the ionic charge (see Fig.\ref{fig1} (b)).
The off-diagonal component are equal to zero. This expression differs from the 
plasma model dielectric function adopted in previous studies in that: 
(i) it has a tensor character with $\epsilon_{xx}=\epsilon_{yy}\neq\epsilon_{zz}$, (ii) through the double sum in the second member it accounts for transitions between lateral sub-bands, whose probability amplitude is expressed by the momentum matrix element between the one electron wavefunctions (\ref{wfc}). It can be easily shown that these 
transitions do not affect the lateral components of the dielectric tensor, which are given by the simple expression of the plasma dielectric function
\begin{equation}
\epsilon_{xx}(\omega)=\epsilon_{yy}(\omega)=1-\frac{\omega_{P}^{2}}{\omega^2}
\label{drudeded}
\end{equation}
because the momentum matrix element for $x$ and $y$ component vanishes.
\section{Results for finite well potential}
In this section we present the results of calculations for finite well potentials. We take $Al$, $Ag$ and $Cs$ corresponding to a radius per electron 
in Bohr units $r_s/a_0$ equal to $2.07$, $3.02$ and $5.62$ respectively, in order to illustrate QSE at different densities and potential depths. The 
value of the potential depth is obtained by summing the metal work function $W$ with the calculated Fermi energy. 
\begin{figure}
\centering
\includegraphics[width=8cm,angle=0]{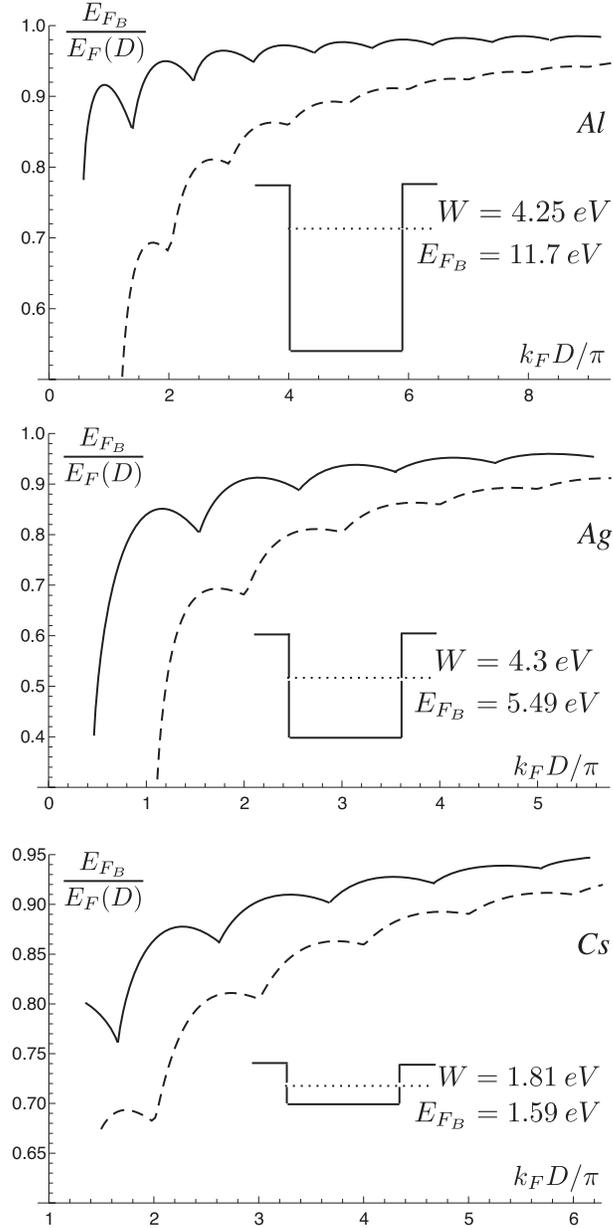}
\caption{\label{fig2} Fermi energy normalized to its bulk value for $Al$, $Ag$ and $Cs$ using the IWM (dashed line) and the FWM (continuous line). The bulk Fermi energies and the work functions have been taken from ref. \cite{ashcroft}.}
\end{figure}
Figure \ref{fig2} reports the calculated Fermi energies as a function of the product between the Fermi wavevector and the film thickness. This allows 
to better point out the oscillations and the cusps arising from the crossing  of the Fermi energy by new subbands upon varying the film size. 
The figure shows the $V_0$ value appropriate to the bulk and to the large $D$ limit. 
We give in the same figure the results obtained by assuming an infinitely deep potential (IWM). The comparison allows to illustrate the effects of the potential softening. In agreement with previously published results \cite{schulter,rogers2,sernelius3} we find that: 
\begin{itemize}
\item  $E_F$ is systematically larger than $E_{F_{B}}$ and goes to the bulk value as $D$ goes to infinity. The difference is more pronounced and the bulk 
limit is achieved at larger size in the low density systems, as it is clearly shown by the comparison between $Cs$ and $Al$ curves;
\item  As expected from the discussion of the previous section, the softening of the potential leads to less pronounced deviations from the bulk 
values. Because of the stronger electron confinement, the IWM has a larger Fermi energy that the FWM;
\item  The cusps correspond to integer values of half of the Fermi wavelength in the IWM case. This feature is only approximately verified for the FWM.
\end{itemize}    
As a consequence of size quantization, the $z$-component of the dielectric tensor is expected to go to a finite value $\epsilon_{zz}(0)$ as the 
frequency goes to zero. This value increases proportionally to $D^2$ in the large size limit \cite{wood}. In Figure \ref{fig3} we plot the quantity  
$\epsilon_{zz}(0)/D^2$ for the three cases under study. 
\begin{figure}
\centering
\includegraphics[width=8cm,angle=0]{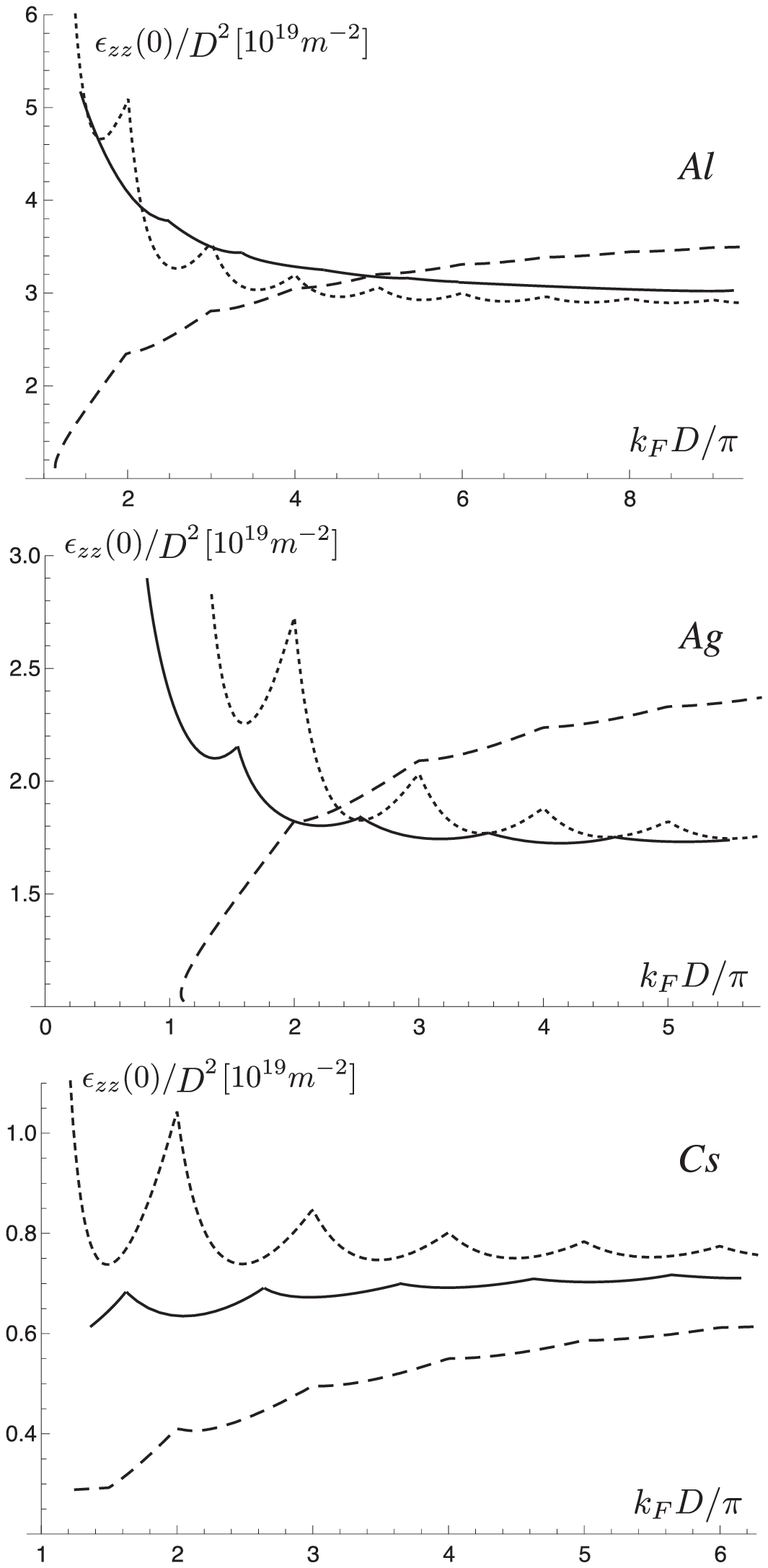}
\caption{\label{fig3} Static value of the $zz$ component of the dielectric tensor for $Al$, $Ag$ and $Cs$ using the IWM (dashed lines), the PBM (dotted lines) and the FWM (continuous lines).}
\end{figure}
Again the results show the cusps due to the filling of new subbands as $D$ increases 
\cite{benassi3}. The asymptotic limit is obtained for $k_{F}D/\pi$ of the order of $5\div 6$ in the three cases. Significant differences appear in the 
large $D$ behaviour when the FWM results are compared with those from the IWM model: $\epsilon_{zz}(0)$ is larger for finite wells of small 
size, while it is smaller at high $D$ values. The convergence to the asymptotic limit is considerably slower for the infinite well, specially in the 
low density metals. This behaviour reflects the differences in the distribution of the eigenvalues of equation (\ref{schrod}). For the infinite well 
there are infinite bound states whose energy scales like $n^2$, see equation (\ref{russi}), and the separation between two successive levels 
increases linearly with $n$. Such behaviour is not present in the FWM, for which equation (\ref{schrod}) has a finite number of eigenvalues 
corresponding to bound states and a continuum spectrum at positive energies. The PBM curve takes values closer to the FWM than to the IWM. This is primarily a consequence of the fact that PBM allows for electron charge spill-out, while in IWM the electron distribution is confined within $D$ i.e. it has the same size of the positive charge.\\
In the following we show the changes in the force caused by the size quantization with respect to the results obtained by using the isotropic continuum 
plasma model, where the $z$-component of the film dielectric tensor is equal to the planar components i.e. is given by equation (\ref{drudeded}). We 
use the symbol $F_Q$ to indicate the force per unit area calculated for the quantized film, while $F_P$ is the force per unit area calculated in the 
isotropic plasma model. To better illustrate the results, in figures \ref{fig4} and \ref{fig5} we plot the quantity
\begin{equation}
\delta_{P}=\frac{F_P-F_Q}{F_P}
\end{equation}
as a function of the separation distance $\ell$ for films of $1$ and $5$ nm thickness respectively. In each figure we display the results for the 
three cases under study and we compare the finite well with the IWM at the same density. This allows us to point out the modifications caused by the 
potential softening.  We also show the curves appropriate to the PBM.
\begin{figure}
\centering
\includegraphics[width=8cm,angle=0]{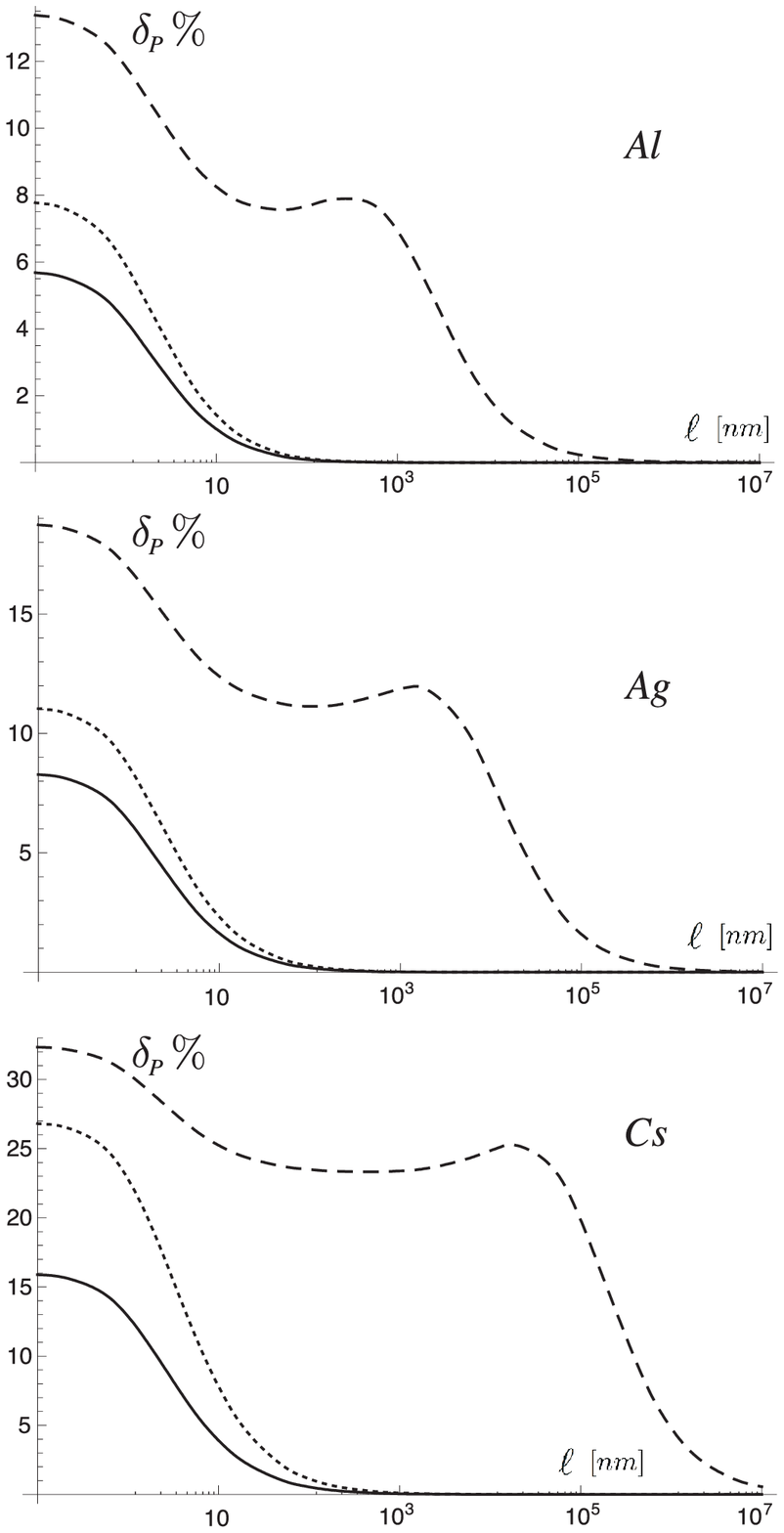}
\caption{\label{fig4} Relative percentual difference for the force between two identical films of thickness $D=1$ nm as a function of the films separation $\ell$, for $Al$, $Ag$ and $Cs$. Dashed lines have been obtained using the PBM, dotted lines have been obtained using the IWM and the continuous lines represents the finite well results.}
\end{figure}
\begin{figure}
\centering
\includegraphics[width=8cm,angle=0]{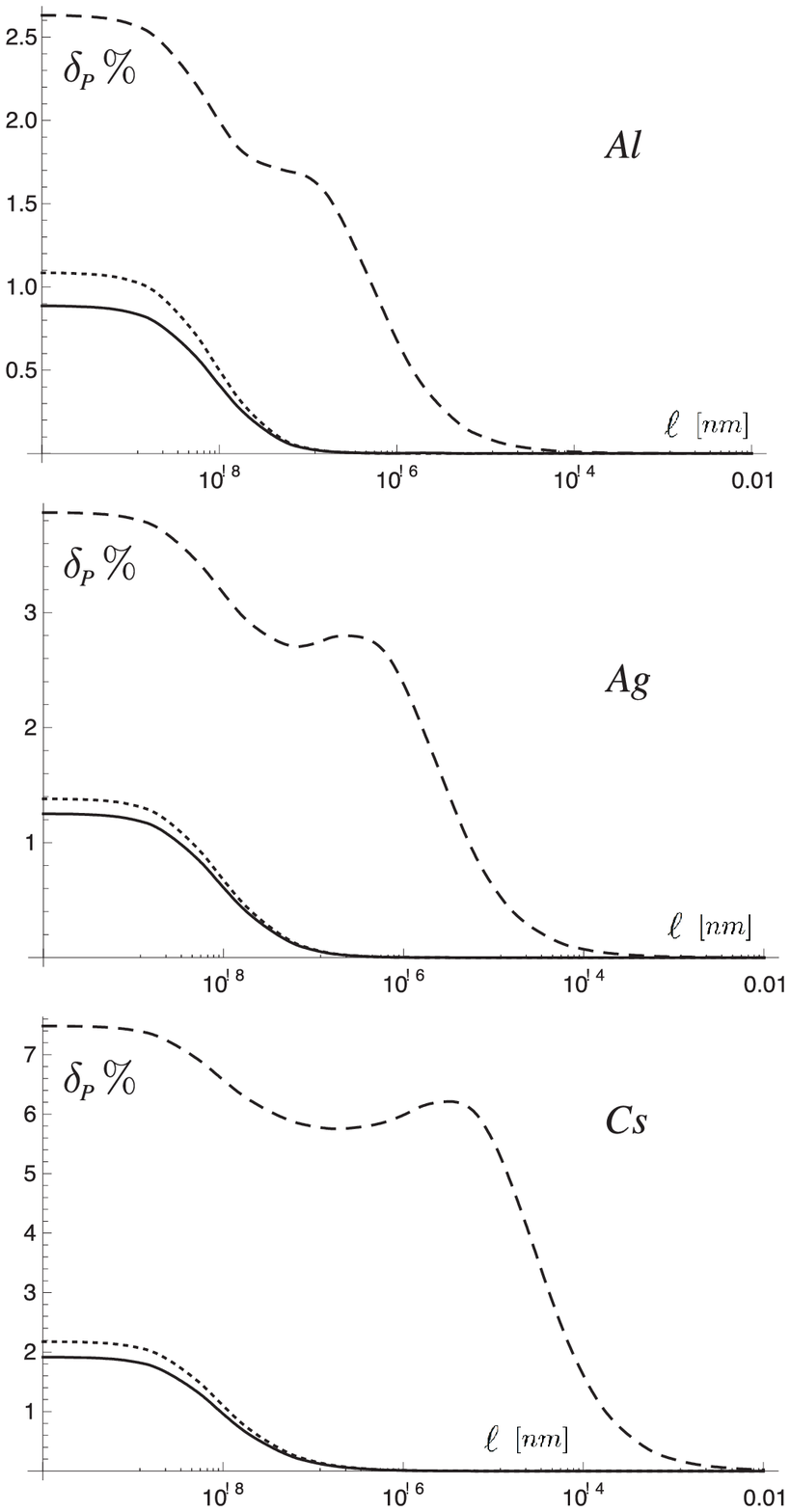}
\caption{\label{fig5} Relative percentual difference for the force between two identical films of thickness $D=5$ nm as a function of the films separation $\ell$, for $Al$, $Ag$ and $Cs$. Dashed lines have been obtained using the PBM, dotted lines have been obtained using the IWM and the continuous lines represents the finite well results.}
\end{figure}
In agreement with our previous findings we observe that 
\begin{itemize}
\item  QSE tend to reduce the intensity of the force;
\item  the reduction is more significant at low density ($Cs$) than at high density ($Al$); 
\item  it may be considerably higher than $10\%$ for $1$ nm thickness and reduce to few per cent at $5$ nm;
\item  it can be appreciable over a distance interval up to $10\div 50$ nm.
\end{itemize}
The most important conclusion that can be drawn from the figures is that the potential shape is important and can lead to a substantial  
modifications of the quantum size effects both at small and at large distances. The models which confine the electronic charge tend to overestimate 
the force reduction induced by size quantization. The curves for the PBM show large force reduction (greater $\delta_P$ values) over a wide interval of distances. 
On passing to the IWM case one notice that the 
removal of the constriction that the Fermi energy be equal to the bulk value, still keeping an infinitely deep potential, leads to smaller $\delta_P$ 
values and to a more rapid decay of the curves at large distances. Reducing the well depth to finite values has a similar effect: it causes a general 
decrease of $\delta_P$ and a narrowing of the distance interval over which QSE are appreciable. 
This also implies that any increase of the confining potential depth at fixed ion density leads to higher $\delta_P$ values and to more significant 
QSE.
The large values taken by $\delta_{P}$ in the PBM case do not arise from the charge confinement only, since, as pointed out before, the constraint on 
the charge distribution is weaker that in the IWM. To a large extent they are a consequence of the plasma frequency normalization caused by the 
decrease in the average electron charge density that it is necessary in order to achieve global charge neutrality \cite{benassi3,czoschke,czoschke2}. 
In the isotropic plasma model one takes $\omega_{P}=\Omega_{P}$. In the PBM this value is obtained only at large film thickness. Neglecting this 
normalization i.e. taking the free electron plasma frequency in the parallel components of the dielectric tensor (but not in $\epsilon_{zz}$) would 
lead to $\delta_{P}$ values  closer to the well potential models.
\section{Intraband absorption effects}
\begin{figure}
\centering
\includegraphics[width=8cm,angle=0]{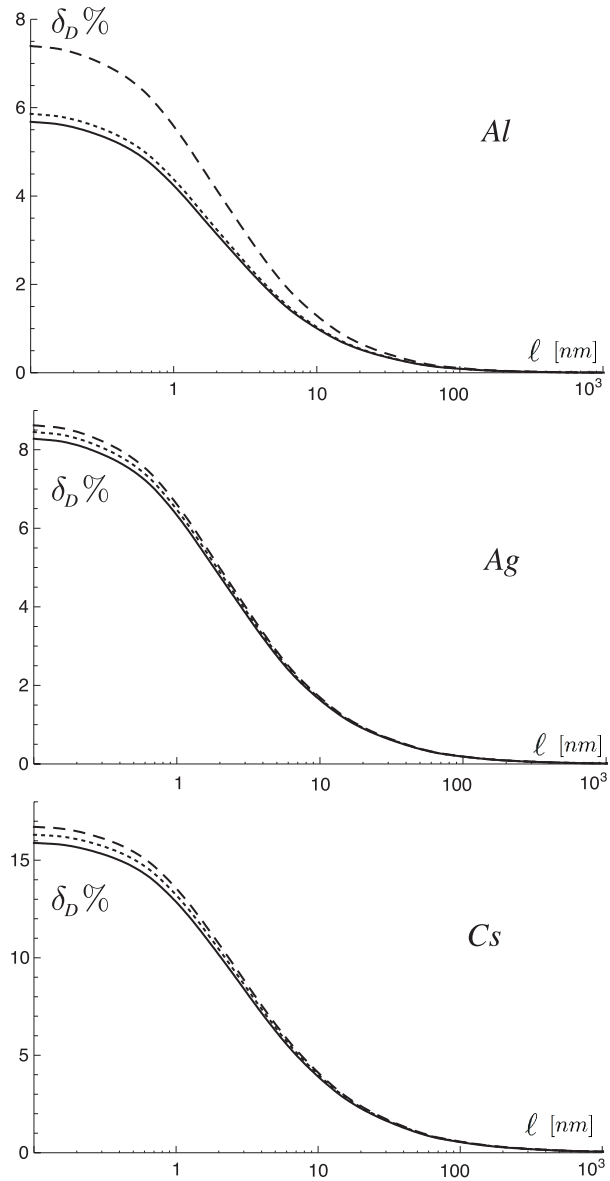}
\caption{\label{fig6} Relative percentual difference for the force between two identical films of thickness $D=1$ nm as a function of the films 
separation $\ell$, for $Al$, $Ag$ and $Cs$. Continuous lines represent the results for $\gamma=0$. Dotted lines have been obtained using 
$\gamma=5\times 10^{13}$ rad/s for $Ag$ and $Cs$ and $\gamma=10^{14}$ rad/s for $Al$. Dashed lines have been obtained using $\gamma=10^{14}$ rad/s 
for $Ag$ and $Cs$ and $\gamma=10^{15}$ rad/s for $Al$.}
\end{figure}

\begin{figure}
\centering
\includegraphics[width=8cm,angle=0]{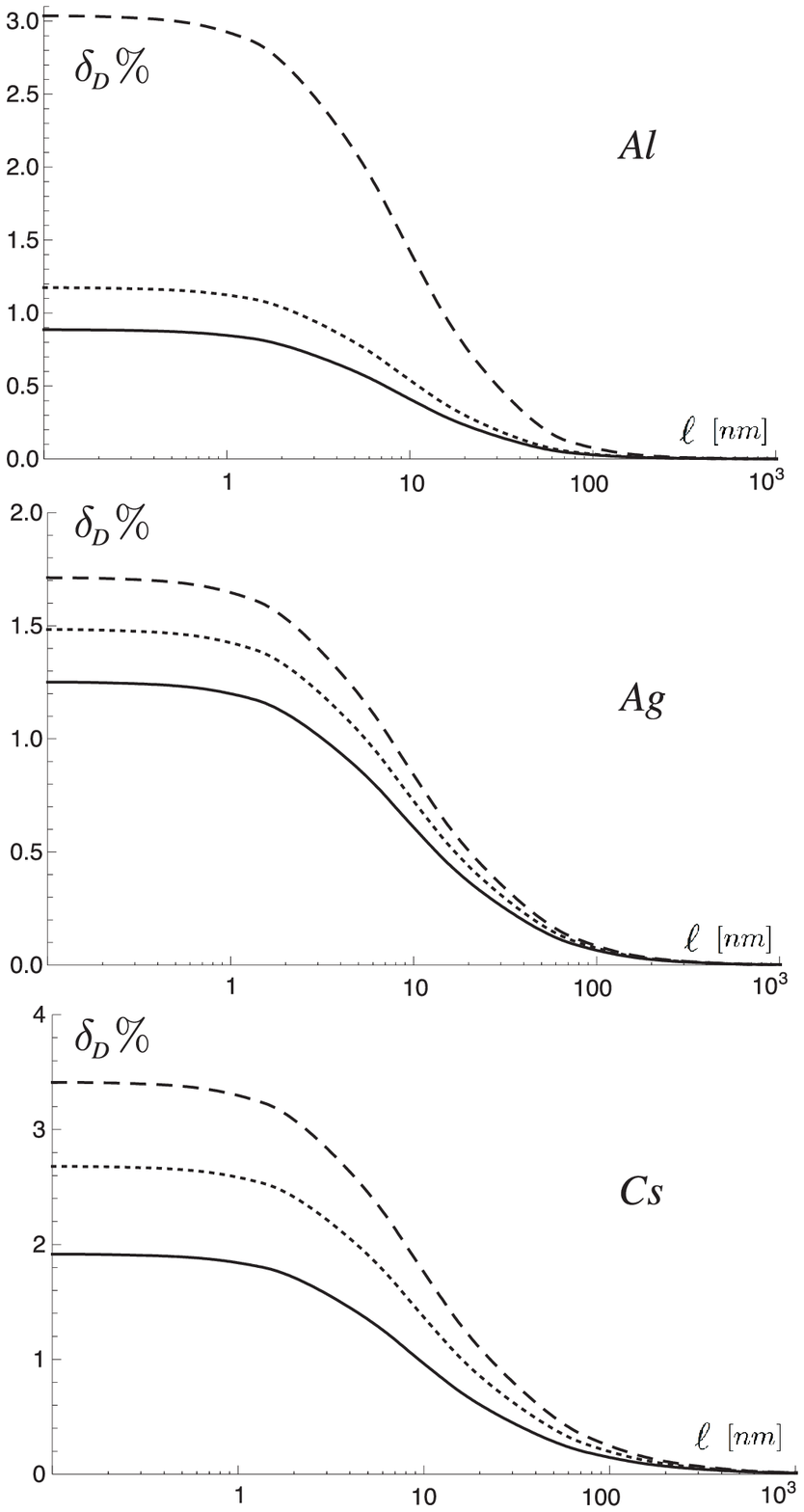}
\caption{\label{fig7} Relative percentual difference for the force between two identical films of thickness $D=5$ nm as a function of the films 
separation $\ell$, for $Al$, $Ag$ and $Cs$. Continuous lines represent the results for $\gamma=0$. Dotted lines have been obtained using 
$\gamma=5\times 10^{13}$ rad/s for $Ag$ and $Cs$ and $\gamma=10^{14}$ rad/s for $Al$. Dashed lines have been obtained using $\gamma=10^{14}$ rad/s 
for $Ag$ and $Cs$ and $\gamma=10^{15}$ rad/s for $Al$.}
\end{figure}

\begin{figure}
\centering
\includegraphics[width=8cm,angle=0]{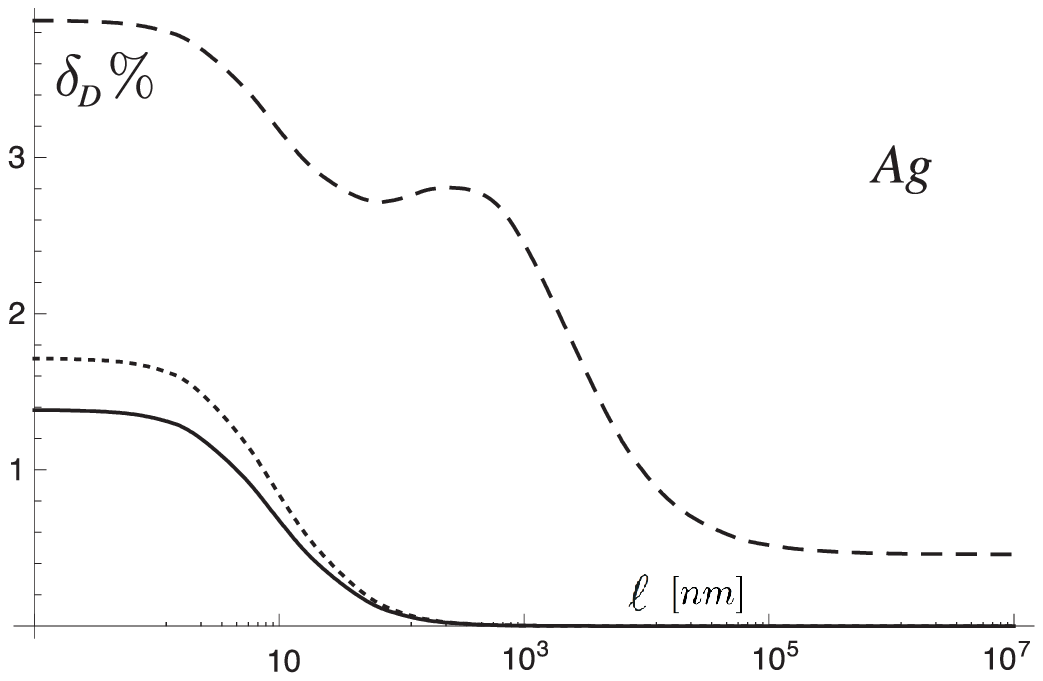}
\caption{\label{fig8} Relative percentual difference for the force between two identical $Ag$ films of thickness $D=5$ nm as a function of the films 
separation $\ell$, $\gamma=10^{14}$ rad/s. Continuous lines represent the FWM, dotted line the IWM and dashed line the PBM.}
\end{figure}

To introduce intraband absorption we have to modify the dielectric tensor in a way that allows to include relaxation time effects in the parallel 
components and to recover  the Drude behaviour in the large $D$ limit. This cannot be done by simply introducing an imaginary part of  the frequency 
$\omega$, since this violates the continuity equation locally \cite{mermin,garik}. The appropriate recipe is to replace into equation 
(\ref{dieltens}) $\omega^2$  with $\omega(\omega+i 2 \pi/\tau)$, where $\tau$ is the relaxation time.
For the parallel components this leads to the Drude dielectric function
\begin{equation}
\epsilon(\omega)=1+\frac{\omega_{P}^2}{\omega(\omega+i \gamma)}
\end{equation}
where $\gamma=2 \pi /\tau$
The results of the calculation of the force per unit area with the modified dielectric tensor are displayed in Figures \ref{fig6} and \ref{fig7} for 
$1$ nm and $5$ nm films. We report the quantity
\begin{equation}
\delta_{D}(\gamma)=\frac{F_D(\gamma)-F_{QD}(\gamma)}{F_D(\gamma)}
\end{equation}
where $F_D$ is the force calculated using the bulk Drude model with a given relaxation time and $F_{QD}$ in the one obtained by the calculation with 
the same relaxation time and with size quantization included. For $\tau\rightarrow\infty$ we recover the plasma model so that 
$\delta_P=\delta_D(0)$.
The calculations have been performed assuming a finite potential well and for two values of  the relaxation frequency $\gamma$. We have taken values 
that approximately correspond to those reported for the metals under consideration \cite{ashcroft}. The figures show a comparison with the curves 
obtained by the plasma model with QSE. 
It is clear from these results that the main effect of the inclusion of intraband absorption is to increase $\delta$ i.e. to increase the difference 
with respect to the calculations with the bulk dielectric function. The smaller is the relaxation time the larger is the reduction of the force. The 
effect is qualitatively the same in the three metals under study, but it depends upon the well depth and the film size.
The influence of the shape of the potential is illustrated in Fig.\ref{fig8} showing the curves of $Ag$ films at a given relaxation frequency for the 
various models. Again it should be noticed that the FWM gives the lower $\delta_D$ values. The PBM results show large $\delta_D$ values over a very 
wide interval of distances. To a large extent this behaviour has to be imputed to the renormalization of the plasma frequency.
 \begin{figure}
\centering
\includegraphics[width=8cm,angle=0]{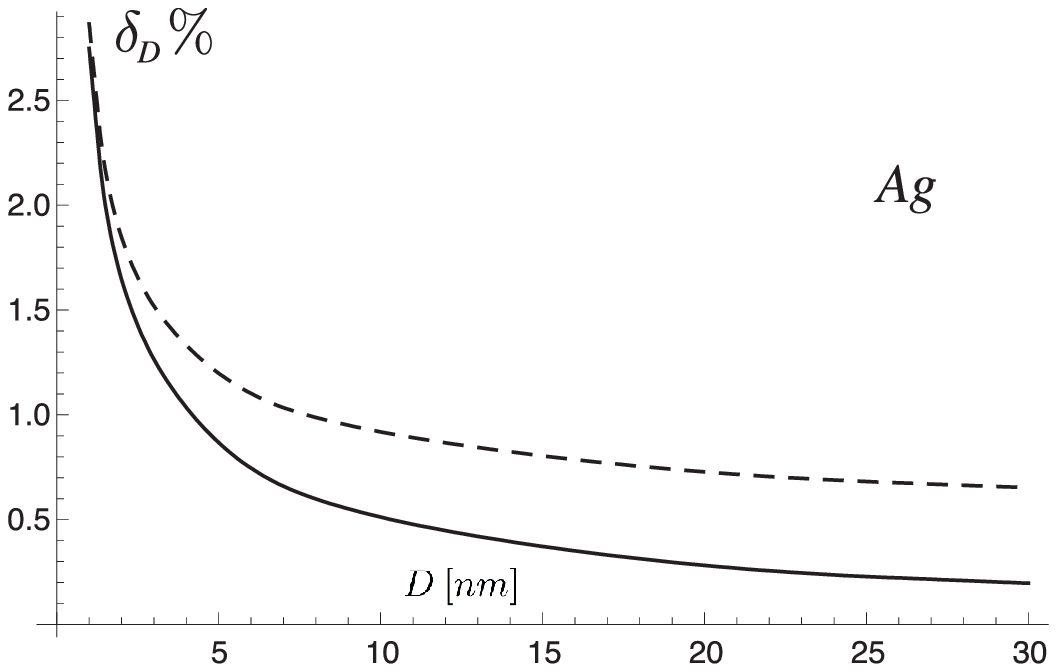}
\caption{\label{fig9} Relative percentual difference for the force between two identical $Ag$ films separated by a distance $\ell=5$ nm as a function of the films thickness $D$. Continuous line represents the result for $\gamma=0$ whereas the dashed line has been obtained with $\gamma=10^{14}$ rad/s. }
\end{figure}
Figure \ref{fig9} shows typical curves of $\delta_D$ as a function of the film thickness for different values of the relaxation frequencies at a 
given separation distance of $5$ nm. As expected $\delta_D$ decreases with $D$, but the slope at large thicknesses ($D$ of the order of $10\div50$ 
nm) depends significantly upon the relaxation frequency. 
\section{Conclusions and lines of development}
We have presented a rather complete set of theoretical results illustrating the possible role of size quantization effects in the electromagnetic 
vacuum force between very thin films and showing how the determination of these effects depends upon the description of the film electronic structure 
and upon the inclusion of intraband effects. However we want to point out that our analysis is still a mere indication of the corrections to the 
simple picture that assumes the same dielectric function for films and bulk solids. Although the basic features of size quantization (confinement of 
the electronic charge, anisotropy of the dielectric tensor, presence of inter-subbands transitions in the dielectric function) are already present in 
the models we have studied, there is room for substantial improvements before an accurate comparison with experimental data, like those obtained by 
Lisanti et al. for $Pd$ films \cite{lisanti}, can be done. A more detailed treatment should include (i) band structure effects, (ii) non-locality of the dielectric response and (iii) non-local treatment of the reflectivity.
A self consistent first principles calculation of the inverse dielectric matrix for a slab of the appropriate size and symmetry, from which a 
macroscopic dielectric function can be derived with band structure and non-local effects included, could provide the appropriate treatment of the 
first two issues \cite{li,sernelius2}. 
Corrections to the Fresnel optics, along the lines indicated by several authors \cite{appel,feibelman2,kempa}, can lead to important modifications 
of the reflectivity even in the case of a free electron gas film. We are currently investigating these matters and the results will be presented 
elsewhere.
Still when comparing theory with experiments for thin films one should consider the fact that measured relaxation frequency turn out to depend upon 
the film size and morphology \cite{jalochowski2,fahsold,sotelo,pribil,hoffmann}. This is largely due to the so called classical size effects arising from the scattering of the electrons at the 
film boundaries. In view of the sensitivity of QSE to the value of the relaxation time, comparison with experimental data may be possible only if 
a realistic estimate of the modifications in the relaxation times caused by surface scattering is available.
\ack AB thanks \emph{CINECA Consorzio Interuniversitario} ({\tt
www.cineca.it}) for funding his Ph.D. fellowship.
\section*{Reference}
\bibliographystyle{unsrt}

\end{document}